\documentclass[a4paper,cite]{article}

\usepackage{latexsym}

\usepackage{graphicx}
\usepackage{color,pst-plot}
\usepackage{pstricks}

\newcommand{\lbl}[1]{\label{eq:#1}}
\newcommand{ \rf}[1]{(\ref{eq:#1})}

\newcommand{\be}{\begin{equation}}
\newcommand{\ee}{\end{equation}}
\newcommand{\bea}{\begin{eqnarray}}
\newcommand{\eea}{\end{eqnarray}}
\newcommand{\setl}{\setlength\arraycolsep{2pt}}

\newcommand{\noi}{\noindent}
\newcommand{\nn}{\nonumber}
\newcommand{\ra}{\rightarrow}
\newcommand{\Ra}{\Rightarrow}

\newcommand{\lesssim}{ {\
\lower-1.2pt\vbox{\hbox{\rlap{$<$}\lower5pt\vbox{\hbox{$\sim$}}}}\ } 
}
\newcommand{\gtrsim}{ {\
\lower-1.2pt\vbox{\hbox{\rlap{$>$}\lower5pt\vbox{\hbox{$\sim$}}}}\ } 
}

\newcommand{\cL}{{\cal L}}

\newcommand{\cO}{{\cal O}}

\newcommand{\Imm}{\mbox{\rm Im}}

\newcommand{\tr}{\mbox{\rm tr}}

\newcommand{\MeV}{\mbox{\rm MeV}}

\newcommand{\annd}{\mbox{\rm and}}

\newcommand{\bfw}{\mbox{\bf w}}
\newcommand{\bfg}{\mbox{\bf g}}

\input epsf

\begin{document}

\begin{titlepage}

\begin{flushright}
CPT-2004/P.014
\end{flushright}

\vspace*{0.2cm}
\begin{center}
{\Large {\bf Rare Kaon Decays Revisited
}}\\[2 cm]

{\bf Samuel Friot}, {\bf David Greynat} {\bf and  Eduardo de Rafael}\\[1cm]

  {\it Centre  de Physique Th{\'e}orique~\footnote{Unit{\'e} Mixte de Recherche (UMR 6207) du CNRS et des Universit{\'e}s Aix Marseille 1, Aix Marseille 2 et sud Toulon-Var, affili{\'e}e {\`a} la FRUMAM}\\
       CNRS-Luminy, Case 907\\
    F-13288 Marseille Cedex 9, France}\\[0.5cm]

\end{center}

\vspace*{1.0cm}

\begin{abstract}
We present an updated discussion of $K\ra\pi \bar{l} l$ decays in a combined framework of chiral perturbation theory and  Large--$N_c$ QCD, which assumes the dominance of a minimal narrow resonance structure in the invariant mass dependence of the $ \bar{l} l$ pair. The proposed picture reproduces very well, both the experimental $K^+\ra\pi^+ e^+ e^-$ decay rate and the invariant $e^+ e^-$ mass spectrum. The predicted Br$(K_S\ra\pi^0 e^+ e^-)$ is, within errors, consistent with the recently reported result from the NA48 collaboration. Predictions for the $K\ra\pi~\mu^{+}\mu^{-}$ modes are also obtained. We find that the resulting interference between the {\it direct} and {\it indirect} CP--violation amplitudes in $K_L\ra\pi^0 e^+ e^-$ is constructive. 
\end{abstract}

\end{titlepage}

\section{Introduction}\lbl{int}

\noi
In the Standard Model, transitions like $K\ra \pi l^+ l^-$, with $l=e,
\mu$, are governed by the interplay of weak non--leptonic and
electromagnetic interactions. To lowest order in the electromagnetic
coupling constant they are expected to proceed, dominantly,
via one--photon exchange. This is certainly the case for the 
$K^{\pm}\ra
\pi^{\pm} l^+ l^-$ and 
$K_{S}\ra \pi^0 l^+ l^-$ decays~\cite{EPR87}. The transition 
$K_{2}^{0}\ra \pi^0\gamma^*\ra \pi^0 l^+ l^-$, 
via one virtual photon, is
however forbidden by CP--invariance. It is then not obvious 
whether the physical decay $K_{L}\ra \pi^0 l^+ l^-$ will still be
dominated by the CP--suppressed
$\gamma^*$--virtual transition or whether a transition via two virtual
photons, which is of higher order in the electromagnetic coupling but
CP--allowed, may dominate~\cite{EPR88}. The possibility of reaching branching
ratios for the mode $K_{L}\ra
\pi^{0}e^{+}e^{-}$  as small as $10^{-12}$ in the near future 
dedicated experiments of the NA48 collaboration at CERN, is a strong motivation for an update of the theoretical
understanding of these modes.

The CP--allowed transition $K_{2}^{0}\ra \pi^{0}\gamma^*\gamma^*\ra
\pi^{0}e^{+}e^{-}$  has been extensively studied in the 
literature (see refs.~\cite{CEP93,DG95} and references therein). We have nothing new to report on this mode. A recent estimate of a conservative upper bound for this transition gives a branching ratio~\cite{BDI03}
\begin{equation}
	{\rm Br}(K_L \ra \pi^0 e^+ e^-)\vert_{\mbox{\rm\tiny CPC}}< 3\times 10^{-12}\,. 
\end{equation}

There are two sources of CP--violation in the transition 
$K_{L}^{0}\ra \pi^0\gamma^*\ra \pi^0 l^+ l^-$. The  {\it direct} source is the  one
induced by the ``electroweak penguin''--like diagrams which generate
the effective local four--quark operators~\cite{GW80} 

\be
\lbl{q11q12}                                                               
Q_{11} =  4 \, (\bar s_L \gamma^\mu d_L)                                
\sum_{l=e,\mu} \, (\bar l_L \gamma_\mu l_L) \qquad\annd \qquad
Q_{12} = 4 \, (\bar s_L \gamma^\mu
d_L)                                      
\, \sum_{l=e,\mu}  (\bar l_R \gamma_\mu l_R) \, 
\ee
\noi
modulated by Wilson coefficients which have an imaginary part induced by
the CP--violation phase of the flavour mixing matrix. The  {\it
indirect} source of CP--violation is the  one induced by the
$K_{1}^{0}$--component of the $K_{L}$ state which brings in the
CP--violation parameter $\epsilon$. The problem in the {\it
indirect} case is, therefore, reduced to the evaluation of the CP--conserving
transition
$K_{1}^{0}\ra
\pi^{0}e^{+}e^{-}$. If the sizes of the two CP--violation sources are comparable, as phenomenological estimates seem to indicate~\cite{EPR88,DG95,DEIP98,BDI03}, the induced branching ratio  becomes, of course, rather sensitive to the interference between the two  {\it direct} and {\it indirect} amplitudes. Arguments in favor of a constructive interference have been recently suggested~\cite{BDI03}.

The analysis of $K\ra \pi\gamma^*\ra \pi l^{+}l^{-}$
decays within the framework of chiral perturbation theory ($\chi$PT) was first made in
refs.~\cite{EPR87,EPR88}. To lowest non trivial order 
in the chiral expansion, the corresponding decay amplitudes
get contributions both from chiral one loop graphs, and from tree level
contributions of local operators of $\cO (p^4)$. In fact, only two
local operators of the $\cO (p^4)$   effective Lagrangian with $\Delta S=1$
contribute to the amplitudes of these decays. With
$\cL_{\mu}(x)$ the $3\times 3$ flavour matrix current field

\be \label{eq:leftc}
\cL_{\mu}(x)\equiv -iF_{0}^2 U(x)^{\dag}D_{\mu}U(x)
\,, 
\ee
where $U(x)$ is the matrix field which collects the Goldstone fields ($\pi$'s,  $K$'s and $\eta$), 
the relevant effective Lagrangian as written in ref.~\cite{EPR87}, is 
\be \label{eq:effl}
\cL_{\rm eff}^{\Delta S=1}(x)\doteq\
-\frac{G_{F}}{\sqrt{2}}\,V_{\rm ud}^{\phantom{\ast}} 
V_{\rm us}^{\ast}
\,{\bfg}_{8}\left\{\tr\left(\lambda\cL_{\mu}\cL^{\mu}
\right) -\,\frac{ie}{F_{0}^2}
\left[{\bfw}_{1}\, {\tr}(Q\lambda\cL_{\mu}\cL_{\nu})+
{\bfw}_{2}\, {\tr}(Q\cL_{\mu}\lambda\cL_{\nu})\right]
\,F^{\mu\nu}\right\}+
\mbox{\rm h.c.}\,.
\ee
\noi
\noi 
Here $D_{\mu}$ is a covariant derivative which, in the
presence of an external
electromagnetic field source $A_{\mu}$ only, reduces to
$D_{\mu}U(x)=\partial_{\mu} U(x)-ie A_{\mu}(x) [Q,U(x)]\,;$ 
$F^{\mu\nu}$is the
electromagnetic field strength tensor; $F_0$ is the
pion decay coupling constant
($F_0\simeq 87$ MeV) 
in the chiral limit;  
$Q$ the electric charge matrix; and $\lambda$ a short--hand notation for
the $SU(3)$ Gell-Mann matrix $(\lambda_{6}-i\lambda_{7})/2$: 
\be
\label{eq:Q} 
Q=\left (
          \begin{array}{ccc}
           2/3 & 0 & 0 \\
           0 & -1/3 & 0 \\
           0 & 0 & -1/3
           \end{array}
          \right )\,, \qquad\qquad
\lambda=\left (
          \begin{array}{ccc}
           0 & 0 & 0 \\
           0 & 0 & 0 \\
           0 & 1 & 0
           \end{array}
          \right )\,.  
\ee 
\noi
The overall constant ${\bfg}_{8}$ is the dominant coupling of non--leptonic weak transitions with $\Delta S=1$ and
$\Delta I=1/2$ to lowest order in the chiral expansion. The factorization of ${\bfg}_{8}$ in the two couplings ${\bfw}_{1}$ and ${\bfw}_{2}$ is, however, a convention.  

For the purposes of this paper, we shall rewrite the effective
Lagrangian in Eq.~\rf{effl} in a more convenient way.
Using the relations
\be
Q\lambda=\lambda Q=-\frac{1}{3}\lambda\qquad\annd\qquad
Q=\hat{Q}-\frac{1}{3}I\quad \left[\hat{Q}=\mbox{\rm
diag.}(1,0,0)\,,\ \  I=\mbox{\rm
diag.}(1,1,1)\right]\,,
\ee 
and inserting the current field decomposition 
\be
\cL_{\mu}(x)=L_{\mu}(x)-eF_{0}^2 A_{\mu}(x)\Delta(x)\,,
\ee
where
\be
L_{\mu}(x)=-iF_{0}^2
U^{\dagger}(x)\partial_{\mu}U(x)\qquad\annd\qquad
\Delta(x)=U^{\dagger}(x)[\hat{Q},U(x)]\,,
\ee
in Eq.~\rf{effl}, results in the Lagrangian

{\setl
\bea \lbl{efflnex}
\cL_{\rm eff}^{\Delta S=1}(x) & \doteq & 
-\frac{G_{F}}{\sqrt{2}}\,V_{\rm ud}^{\phantom{\ast}} 
V_{\rm us}^{\ast}
\,{\bfg}_{8}\left\{\tr\left(\lambda L_{\mu}L^{\mu}
\right)-eF_{0}^2
A_{\mu}\tr[\lambda(L^{\mu}\Delta+\Delta L^{\mu})]\right. \nn
\\ 
 & + & \frac{ie}{3F_{0}^2}F^{\mu\nu}\left.\left[({\bfw}_{1}-
{\bfw}_{2})\,\ \tr\left(\lambda L_{\mu}L_{\nu}\right)
+3\ 
{\bfw}_{2}\,\ \tr(\lambda L_{\mu}\hat{Q}L_{\nu})\right]\right\}
+\mbox{\rm h.c.}
\eea}

The $Q_{11}$ and $Q_{12}$ operators in Eq.~\rf{q11q12} are
proportional to the quark current density $(\bar{s}_{L}\gamma^{\mu}d_{L})$ and, therefore, their effective
chiral realization can be
directly obtained from the strong chiral Lagrangian [
$(\bar{s}_{L}\gamma^{\mu}d_{L})\Ra 
(\cL^{\mu})_{23}$ to $\cO(p)$].
Using the equations of motion for the leptonic fields $\partial^{\mu}F_{\mu\nu}=e\bar{l}\gamma_{\mu}l$, and doing a partial integration in the action, it follows that the effect of the electroweak penguin operators induces a contribution to the coupling constant ${\bf \tilde{w}}$ only; more precisely
\begin{equation}
	 {\bfg}_{8}\left({\bf \tilde{w}}={\bfw}_{1}-{\bfw}_{2}\right)\bigg\vert_{Q_{11},Q_{12}}= 
	 \frac{3}{4\pi\alpha}\left[C_{11}(\mu^2)+C_{12}(\mu^2) \right]\,,
\end{equation}
where $C_{11}(\mu^2)$ and $C_{12}(\mu^2)$ are the Wilson coefficients of the $Q_{11}$ and $Q_{12}$ operators. There is a resulting $\mu$--scale dependence in the real part of the Wilson coefficient $C_{11}+C_{12}$  due to an incomplete cancellation of the GIM--mechanism because, in the short--distance evaluation, the $u$--quark has not been integrated out. This $\mu$--dependence should be canceled when doing the matching with the long--distance evaluation of the weak matrix elements of the other four--quark operators; in particular, with the contribution from the unfactorized pattern of the $Q_2$ operator in the presence of electromagnetism. It is in principle possible, though not straightforward,  to evaluate  the $\tilde {\bf w}$ and $\bfw_2$ couplings within the framework of Large--$N_c$ QCD, in much the same way as other low--energy constants have been recently determined (see e.g. ref.~\cite{HPdeR03} and references therein). While awaiting the results of this program, we propose in this letter a more phenomenological approach. Here we shall discuss the determination of the couplings  $\tilde {\bf w}$ and $\bfw_2$ using theoretical arguments inspired from Large--$N_c$ considerations, combined with some of the experimental results which are already available at present. As we shall see, our conclusions have interesting implications for the CP--violating contribution to the  $K_L\ra \pi^0 e^+ e^-$ mode.  

\section{$K\ra \pi l\bar{l}$ Decays to $O(p^4)$ in the Chiral Expansion}
\setcounter{equation}{0}

As discussed in ref.~\cite{EPR87}, at $\cO(p^4)$ in the chiral
expansion, besides the contributions from the ${\bfw}_{1}$ and ${\bfw}_{2}$ terms in Eq.~\rf{effl} there also appears a tree level contribution to the $K^{+}\ra
\pi^{+}e^{+}e^{-}$ amplitude induced by the combination of the lowest $\cO(p^2)$ weak $\Delta
S=1$ Lagrangian (the first term in Eq.~\rf{effl}) with the  $\mbox{\bf
L}_{9}$--coupling of the 
$\cO(p^4)$ chiral Lagrangian which describes strong interactions in the
presence of electromagnetism~\cite{GL85}:
\be\lbl{l9}
\cL_{\rm em}^{(4)}(x)\doteq -ie\mbox{\bf L}_{9}F^{\mu\nu}(x)\tr\left\{ 
Q\ D_{\mu}U(x)D_{\nu}U^{\dagger}(x)+
Q\ D_{\mu}U(x)^{\dagger}D_{\nu}U(x)
\right\}\,.
\ee
 In full generality, one can then predict the
$K^{+}\ra \pi^{+} l^{+}l^{-}$ decay rates ($l=e,\mu$) as a function of
the scale--invariant combination of coupling constants
\be
\label{eq:kppiplplmw}   {\bfw}_{+}= -\frac{1}{3}(4\pi)^{2}
\,[{\bfw}_{1}-{\bfw}_{2}+3({\bfw}_{2}-4\mbox{\bf L}_{9})]
-\frac{1}{6}\log\frac{M_{K}^2 m_{\pi}^{2}}{\nu^4}\,,
\ee
\noi   where ${\bfw}_{1}$, ${\bfw}_{2}$ and $\mbox{\bf L}_{9}$ are
renormalized couplings
at the scale $\nu$. 
The coupling constant
$\mbox{\bf L}_{9}$ can be determined from the electromagnetic mean
squared radius of the pion~\cite{BEG94}: $\mbox{\bf L}_{9}(M_{\rho})=
(6.9\pm 0.7)\times 10^{-3}$. The combination of constants ${\bfw}_{2}-4\mbox{\bf L}_{9}$ is in fact scale independent. 
To that order in the chiral expansion, the predicted decay rate $\Gamma(K^+\ra\pi^+ e^+ e^-)$ as a function of ${\bfw}_{+}$ describes a parabola. The intersection of this parabola with the experimental decay rate obtained from the branching ratio~\cite{K+dacrate}
\begin{equation}\lbl{K+rate}
{\rm Br}(K^+\ra\pi^+ e^+ e^-)=(2.88\pm0.13)\times 10^{-7}\,,	
\end{equation}
gives the two phenomenological  solutions (for a value of the overall constant ${\bf g}_8=3.3$):
\begin{equation}\lbl{solsK+}
	{\bfw}_{+}=1.69\pm 0.03\quad\annd\quad {\bfw}_{+}=-1.10\pm 0.03\,.
\end{equation}
Unfortunately, this twofold determination of the
constant 
${\bfw}_{+}$ does not help to predict the $K_{S}\ra \pi^{0}e^{+}e^{-}$
decay rate. This is due to the fact that, to the same order in the
chiral expansion, this transition amplitude brings in
another scale--invariant combination of constants:
\be \lbl{eq:ws}
{\bfw}_{s}= -\frac{1}{3}(4\pi)^{2}\,[{\bfw}_{1}-{\bfw}_{2}]
-\frac{1}{3}\log\frac{M_{K}^{2}}{\nu^2}\,.
\ee
The predicted decay rate $\Gamma(K_S \ra\pi^0 e^+ e^-)$ as a function of ${\bfw}_{s}$ is also a parabola. From the recent result on this mode, reported by the NA48 collaboration at CERN~\cite{NA48}:
 \begin{equation}\lbl{K0rate}
	{\rm Br}\left(K_S \rightarrow \pi^0 e^+ e^-\right)= \left[5.8^{+2.8}_{-2.3} (\rm stat.) \pm 0.8 (\rm syst.) \right] \times 10^{-9}\,,
\end{equation} 
one obtains the two solutions for ${\bfw}_{s}$
\begin{equation}\lbl{solsK0}
{\bfw}_{s}= 2.56 ^{+0.50}_{-0.53} \quad \annd \quad 	{\bfw}_{s}= -1.90 ^{+0.53}_{-0.50}\,\,.
\end{equation}

At the same $\cO(p^4)$ in the chiral expansion, the branching ratio for the $K_L\ra \pi^0 e^+ e^-$ transition induced by CP--violation  reads as follows

{\setl
\bea\lbl{KSCPV}
	\lefteqn{{\rm Br}\left(K_L \rightarrow \pi^0 e^+ e^-\right)\vert_{\rm\tiny CPV}=} \nn \\
	& & \left[(2.4\pm 0.2) \left(\frac{{\rm Im}\lambda_t}{10^{-4}}\right)^{2}+(3.9\pm 0.1)\,\left(\frac{1}{3}-{\bfw}_s\right)^2+ (3.1\pm 0.2)\, \frac{{\rm Im}\lambda_t}{10^{-4}}\left(\frac{1}{3}-{\bfw}_s\right)\right]\times 10^{-12}\,.	
\eea

\noi
Here, the first term is the one induced by the {\it direct} source, the second one by the {\it indirect} source and the third one the {\it interference} term. With~\cite{Baetal03} $\Imm\lambda_t= (1.36\pm 0.12)\times 10^{-4}$, the interference is constructive for the negative solution in Eq.~\rf{solsK0}. 

The four solutions obtained in Eqs.~\rf{solsK+} and \rf{solsK0}, define four different straight lines in the plane of the coupling constants ${\bfw}_{2}-4{\bf L}_9$ and ${\bf \tilde{w}}\ (={\bfw}_{1}-{\bfw}_{2})$, as illustrated in Fig.~1 below.  We next want to discuss which of these four solutions, if any, may be favored by theoretical arguments.

\begin{figure}[h]

\begin{center}
\includegraphics[width=0.9\textwidth]{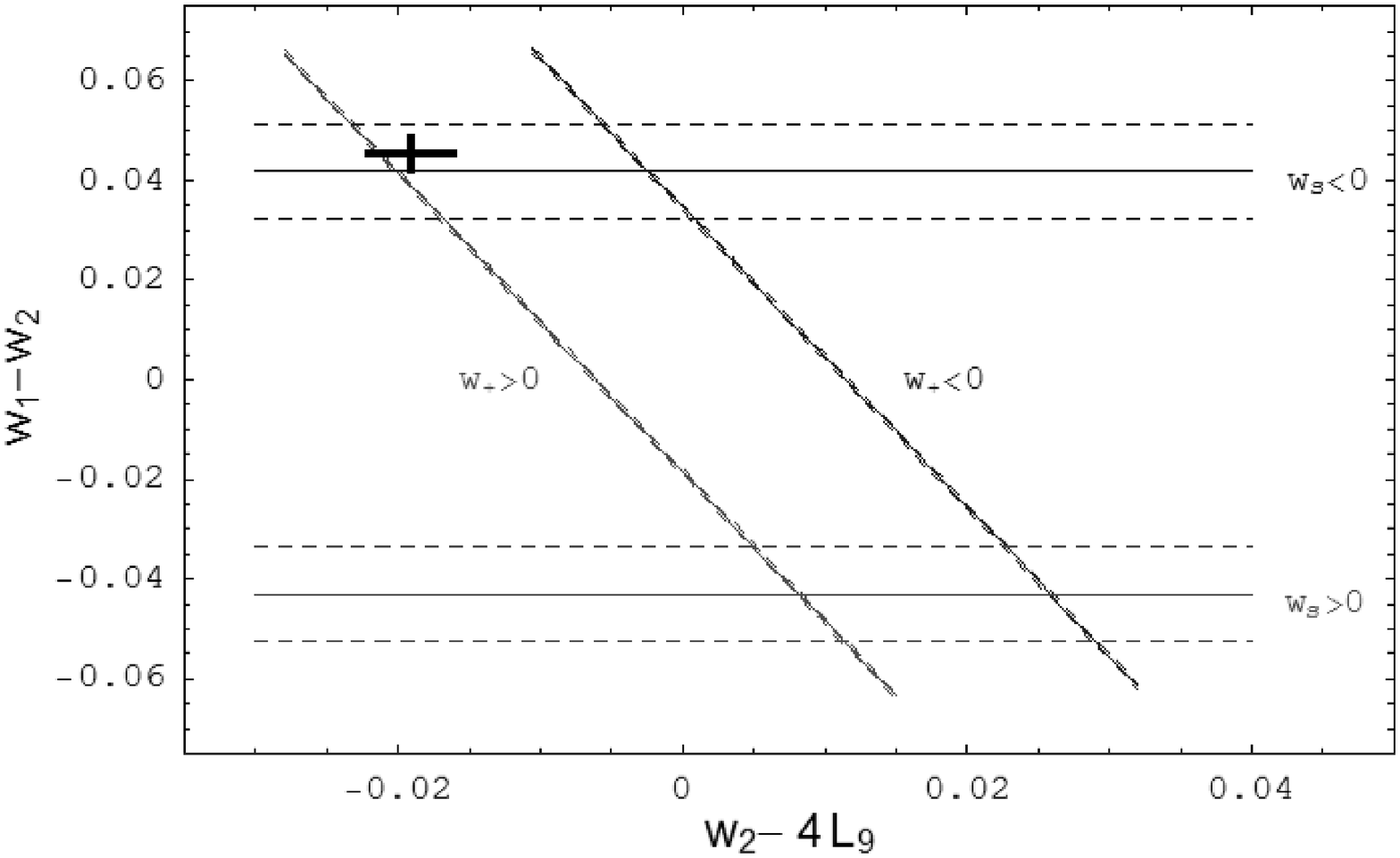}
\end{center}

{\bf Fig.~1} {\it The four intersections in this figure define the possible values of the couplings which, at $\cO(p^4)$ in the chiral expansion, are compatible with the experimental input of Eqs.~\rf{K+rate} and \rf{K0rate}.  The couplings ${\bfw}_{1}$, ${\bfw}_{2}$, and ${\bf L}_9$ have been  fixed at the $\nu=M_{\rho}$ scale and correspond to the value ${\bfg}_8=3.3$. The cross in this figure  corresponds to the values in Eqs.~\rf{wtildepred} and \rf{w24L9pred} discussed in the text. }

\end{figure}

\section{Theoretical Considerations}
\setcounter{equation}{0}

\subsection{The Octet Dominance Hypothesis}

In ref.~\cite{EPR87}, it was suggested that  the couplings ${\bfw}_1$ and ${\bfw}_2$ may satisfy the same symmetry properties as the chiral logarithms generated by the one loop calculation. This selects the octet channel in the transition amplitudes as the only possible channel and leads to the relation 
\begin{equation}\lbl{odh}
	{\bfw}_2=4{\bf L}_9\,,\qquad {\mbox{\rm\footnotesize Octet Dominance Hypothesis (ODH)}}\,.
\end{equation}

We now want to show how this {\it hypothesis} can in fact be justified within a simple dynamical framework of resonance dominance, rooted in Large--$N_c$ QCD. For that, let us examine the field content of the Lagrangian in Eq.~\rf{efflnex}.
For
processes with at most one pion in the final state, it is sufficient
to restrict $\Delta$ and $L_{\mu}$ to their minimum of
one Goldstone field component:
\be
\Delta=-i\frac{\sqrt{2}}{F_{0}}\ [\Phi,\hat{Q}]+\cdots\,,\quad\annd\quad
L_{\mu}=\sqrt{2}F_{0}\ \partial_{\mu}\Phi+\cdots\,,
\ee
with the result (using partial integration in the term proportional to
$ie\,{\bf g}_{8}  {\bfw}_{2}$)

{\setl
\bea \lbl{efflfield}
\cL_{\rm eff}^{\Delta S=1}(x) & \doteq & 
-\frac{G_{F}}{\sqrt{2}}\,V_{\rm ud}^{\phantom{\ast}} 
V_{\rm us}^{\ast}
\,{\bfg}_{8}\left\{2F_{0}^2\  \tr\left(\lambda \partial_{\mu}\Phi
\partial^{\mu}\Phi
\right)+ie\,2F_{0}^2 
A^{\mu}\ \tr[\lambda(\Phi\hat{Q}\partial_{\mu}\Phi-
\partial_{\mu}\Phi\hat{Q}\Phi)]\right.\nn
\\ 
 & &  \ \ \ \ \ \ \ \ \ \ \ \ \ \ \ \ \ \ \ \ \ \ \ \ \ -ie\,{\bfw}_{2}\   \partial_{\nu}F^{\nu\mu}\  
\tr[\lambda(\Phi\hat{Q}\partial_{\mu}\Phi-
\partial_{\mu}\Phi\hat{Q}\Phi)] \nn \\
& & \left. +ie\,\frac{2}{3}{\bf \tilde{w}}\, F^{\mu\nu}\  \tr\left(\lambda \partial_{\mu}\Phi
\partial_{\nu}\Phi
\right)\right\}
 +\mbox{\rm h.c.}
\eea}

\noi
showing that the two--field content which in the term modulated by $ {\bfw}_{2}$ couples to
$\partial_{\nu}F^{\nu\mu}$  is
exactly the same as the one which couples to the gauge
field $A^{\mu}$  in the lowest 
$\cO(p^2)$ Lagrangian. As explained in ref.~\cite{EPR87}, the
contribution to $K^+\ra \pi^+ \gamma$(virtual) from this  $\cO(p^2)$
term, cancels with the one resulting  from the combination of the first term
in Eq.~\rf{efflfield} with the lowest order hadronic electromagnetic
interaction,
in the presence of mass terms for the Goldstone fields. This
cancellation is expected because of the mismatch between the minimum
number of powers of external momenta required by gauge invariance and
the powers of momenta that the lowest order effective chiral Lagrangian
can provide. As we shall next explain, it is the reflect of the
dynamics of this cancellation which, to a first approximation, is also at
the origin of  the relation
$\bfw_2=4\mbox{\bf L}_{9}$.
	
With two explicit Goldstone fields, the hadronic electromagnetic
interaction in the presence of the term in Eq.~\rf{l9} reads as
follows
\be
\cL_{\rm em}(x)=-ie\left( A^{\mu}-\frac{2{\bf L}_{9}}{F_0^2} 
\partial_{\nu} F^{\nu\mu}\right)\
\tr(\hat{Q}\Phi\stackrel{\leftrightarrow}{\partial_{\mu}}\Phi)+\cdots\,.
\ee
The net effect of the ${\bf L}_{9}$--coupling is to provide the
slope of an electromagnetic form factor to the charged Goldstone bosons.
In momentum space this results in a change from the lowest order 
point like coupling to
\be
1\Ra 1-\frac{2{\bf L}_{9}}{F_{0}^2}Q^2\,.
\ee 
In the {\it minimal hadronic approximation} (MHA) to Large--$N_c$ QCD~\cite{PPdeR98},
the form factor in question is saturated by the lowest order pole
i.e. the $\rho(770)$~:
\be
1\Ra \frac{M_{\rho}^2}{M_{\rho}^2+Q^2}\,,\quad{\mbox{\rm which implies}}\quad
{\bf L}_{9}=\frac{F_{0}^2}{2M_{\rho}^2}\,.
\ee
It is well known~\cite{EGPdeR89,EGLPdeR89} that this reproduces the observed slope rather well.

By the same argument, the term proportional to ${\bfw}_{2}$ in Eq.~\rf{efflfield} provides the slope of
the lowest order electroweak coupling of two Goldstone bosons:
\begin{equation}
	\cL_{\rm ew}(x)=-ie\,\frac{G_{F}}{\sqrt{2}}\,V_{\rm ud}^{\phantom{\ast}} 
V_{\rm us}^{\ast}
\,{\bfg}_{8}\ 2F_0^2\left(A^{\mu}-\frac{{\bfw}_{2}}{2F_0^2}\partial_{\nu}F^{\nu\mu}\right) 
\ \tr[\lambda(\Phi\hat{Q}\partial_{\mu}\Phi-
\partial_{\mu}\Phi\hat{Q}\Phi)]+\cdots\,.
\end{equation}
In momentum space this results in a change from the lowest order point
like coupling to
\be
1\Ra 1-\frac{{\bf w}_{2}}{2F_{0}^2}Q^2\,.
\ee
Here, however, the underlying $\Delta S=1$ form factor structure in the same MHA as applied to ${\bf L}_9$, can have contributions both from the $\rho$ and the $K^*(892)$~: 
\begin{equation}
	1\Ra \frac{\alpha M_{\rho}^2}{M_{\rho}^2+Q^2}+
	\frac{\beta M_{K^*}^2}{M_{K^*}^2+Q^2}
	\,, \quad{\mbox{\rm with}}\quad \alpha+\beta=1\,,
\end{equation}
because at $Q^2\ra 0$ the form factor is normalized to one by gauge invariance. 
This fixes the slope to 
\begin{equation}
 \frac{{\bf w}_2}{2F_0^2} =\left(\frac{\alpha}{M_{\rho}^2}+\frac{ \beta}{M_{K^*}^2}\right)\,.
\end{equation}
If, furthermore, one assumes the chiral limit where $M_\rho =M_{K^*}$, there follows then the ODH relation in Eq.~\rf{odh};
a result which, as can be seen in Fig.~1, favors the solution where both ${\bf w}_{+}$ and ${\bf w}_{s}$ are negative, and the interference term in Eq.~\rf{KSCPV} is then constructive.

\subsection{Beyond the $\cO(p^4)$ in $\chi$PT}

A rather detailed measurement of the $e^+ e^-$ invariant mass spectrum in  $K^+\ra \pi^+ e^+ e^-$ decays was reported a few years ago in ref.~\cite{zeller}. The observed spectrum confirmed an earlier result~\cite{Alliegroetal92} which had already claimed that a  parameterization in terms of only ${\bf w}_{+}$ cannot accommodate both the rate and the spectrum of this decay mode. It is this observation which prompted the phenomenological analyses reported in refs.~\cite{DEIP98,BDI03}. Here, we want to show that it is possible to understand the observed spectrum within a simple MHA picture of Large--$N_c$ QCD which goes beyond the $\cO(p^4)$ framework of $\chi$PT but, contrary to the proposals in refs.~\cite{DEIP98,BDI03}, it does not enlarge the number of free parameters.  

We recall that, in full generality~\cite{EPR87}, the $K^+\ra \pi^+ e^+ e^-$ differential decay rate depends only on one form factor $\hat \phi(z)$:
\be\lbl{fullgen}
\frac{d\Gamma}{dz} = \frac{G_8^2\alpha^2M_K^5}{12\pi (4\pi)^4} \lambda^{3/2}\left(1,z,r_\pi^2 \right) \sqrt{1-4\frac{r_\ell^2}{z}} \left(1+2 \frac{r_\ell^2}{z} \right) \left| \hat \phi(z) \right|^2\,,
\ee 
where $q^2=z\,M_{K}^2$ is the invariant mass squared of the $e^+ e^-$ pair, and
\be
G_8=\frac{G_{F}}{\sqrt{2}}\,V_{\rm ud}^{\phantom{\ast}} 
V_{\rm us}^{\ast}\,\bfg_8\,,\quad r_\pi=\frac{m_\pi}{M_K}\,, \quad r_\ell=\frac{m_\ell}{M_K}\,.
\ee 
The relation between $\hat \phi(z)$ and the form factor plotted in Fig.~5 of ref.~\cite{zeller}, which we reproduce here in our Fig.~2 below for $\left|f_V(z)\right|^2$, is
\begin{equation}\lbl{relff}
\label{correspondance}
\left|f_V(z)\right|^2= \left|\frac{G_8}{G_F}  \hat \phi(z)\right|^2 \,.
\end{equation} 

\begin{figure}[h]

\begin{center}
\includegraphics[width=0.9\textwidth]{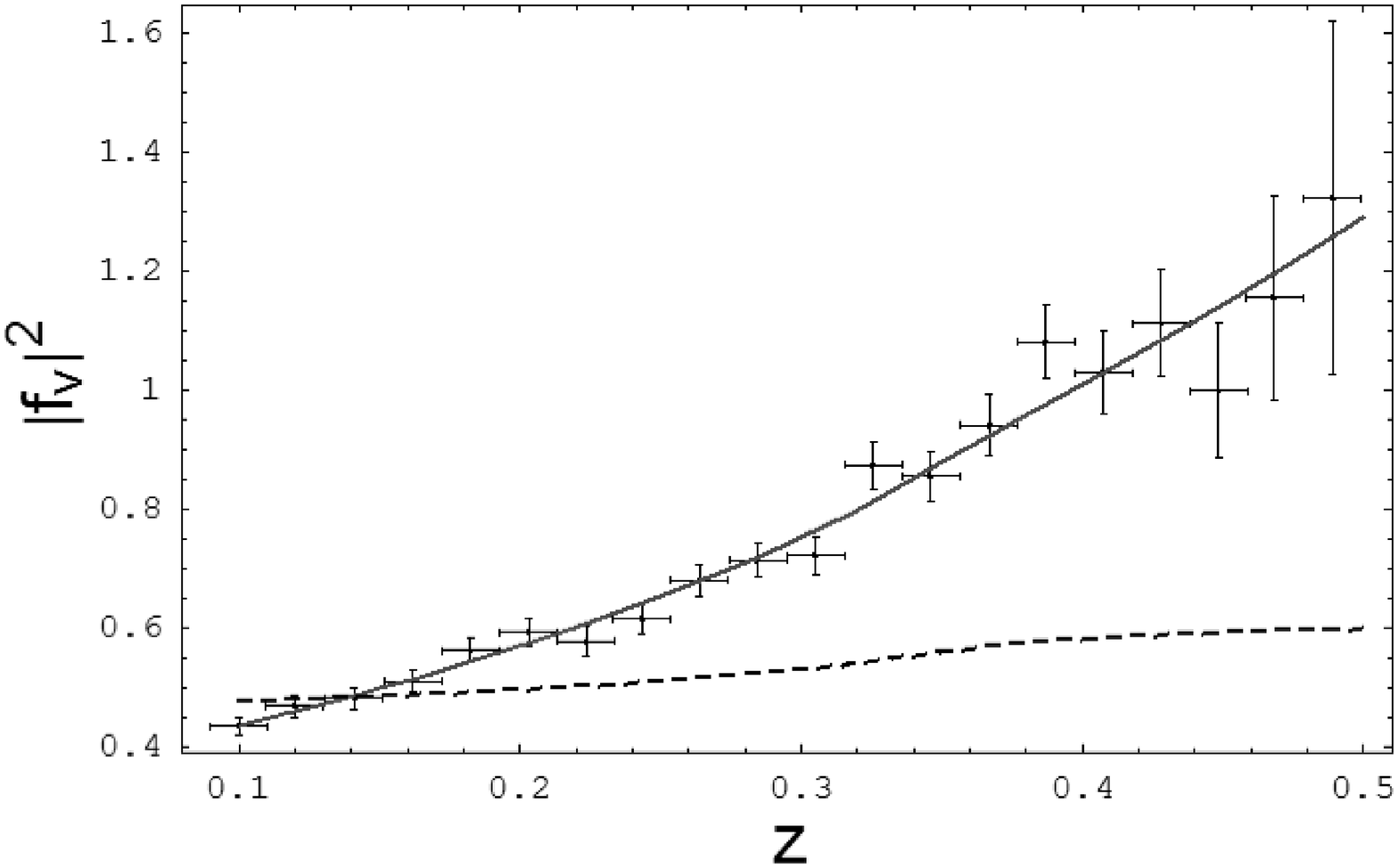}
\end{center}

{\bf Fig.~2} {\it Plot of the form factor $\left|f_V(z)\right|^2$ defined by Eqs.~\rf{fullgen} and \rf{relff} versus the invariant mass squared of the $e^+ e^-$ pair normalized to $M_{K}^2$. The crosses are the experimental points of ref.~\cite{zeller}; the dotted curve is the leading $\cO(p^4)$ prediction, using the positive solution for $\bfw_+$ in Eq.~\rf{solsK+}; the continuous line is the fit to the improved form factor in Eq.~\rf{fff} below.}

\end{figure}
The $\cO(p^4)$ form factor  calculated in ref.~\cite{EPR87} is 
\be
\left|\hat \phi(z)\right|^2= \left| {\bf w}_+ + \phi_K(z) + \phi_\pi(z)\right|^2\,,
\ee
with  the chiral loop functions 
\begin{equation}
\phi_K(z) = -\frac{4}{3} \frac{1}{z} + \frac{5}{18} + \frac{1}{3} \left( \frac{4}{z}-1\right)^{\frac{3}{2}} \arctan \left( \frac{1}{\sqrt{\frac{4}{z}-1}} \right)\quad\annd\quad 
\phi_\pi(z) = \phi\left(z \frac{M_K^2}{m_\pi^2}\right) \,.
\end{equation}
The experimental form factor favors the positive solution in Eq.~\rf{solsK+}, but the predicted $\cO(p^4)$ form factor, the dotted curve in Fig.~2, lies well below the experimental points for $z\gtrsim 0.2$.

Following the ideas  developed in the previous subsection, we propose a very simple generalization of the $\cO(p^4)$ form factor. We keep the lowest order chiral loop contribution as the leading manifestation of the Goldstone dynamics, but replace the local couplings $\bfw_2 -4{\bf L}_9$ and ${\tilde{\bf w}}=\bfw_1-\bfw_2$ in ${\bf w}_{+}$ by the minimal resonance structure  which can generate them in the $z$--channel. For $\bfw_2 -4{\bf L}_9$ this amounts to the replacement:

{\setl
\bea\lbl{w2L9m}
	\bfw_2-4{\bf L}_9 & \Ra & \frac{2F_{\pi}^2}{M_{\rho}^2}\left(\alpha\frac{M_{\rho}^2}{M_{\rho}^2-M_{K}^2 z}
	+\beta\frac{M_{\rho}^2}{M_{K^*}^2}\frac{M_{K^*}^2}{M_{K^*}^2-M_{K}^2 z}-\frac{M_{\rho}^2}{M_{\rho}^2-M_{K}^2 z} \right) \nn \\
	 & = & 	 2F_{\pi}^2\beta \frac{M_{\rho}^2-M_{K^*}^2}{(M_{\rho}^2-M_{K}^2 z)(M_{K^*}^2-M_{K}^2 z)}\,;
\eea}

\noi
while for ${\tilde{\bf w}}$ it simply amounts to the modulating factor:
\begin{equation}\lbl{wtildem}
	{\tilde{\bf w}}\Ra {\tilde{\bf w}}\frac{M_{\rho}^2}{M_{\rho}^2-M_{K}^2 z}\,.
\end{equation}
Notice that in the chiral limit where $M_{\rho}=M_{K^*}$, $F_{\pi}\ra F_0$,  and when $z\ra 0$, we recover the usual $\cO(p^4)$ couplings with the ODH constraint $\bfw_2=4{\bf L}_9$. In our picture, the deviation from this constraint is due to explicit breaking, induced by the strange quark mass, and results in an effective
\begin{equation}
	\bfw_2 -4{\bf L}_9=-\frac{2F_{\pi}^2}{M_{\rho}^2}\beta\left(1- \frac{M_{\rho}^2}{ M_{K^*}^2}\right)\,.
\end{equation}
 
More explicitly, the form factor we propose is 

{\setl
\bea\lbl{fff}
f_V(z) & = & \frac{G_8}{G_F} \left\{  \frac{(4\pi)^2}{3} \left[{\tilde{\bf w}}\frac{M_{\rho}^2}{M_{\rho}^2-M_{K}^2 z}  + 6   F_{\pi}^2 \beta \frac{M_\rho^2 - M_{K^*}^2}{\left(M_\rho^2 - M_K^2 z\right)\left(M_{K^*}^2-M_K^2 z\right)}\right] \right.  \nn \\
 &  & \left.    + \frac{1}{6} \ln \left(\frac{M_K^2 m_\pi^2}{M_\rho^4}\right) +\frac{1}{3} - \frac{1}{60}z - \chi(z) \right\}\,,
\eea}

\noi
where the first line incorporates the modifications in Eqs.~\rf{w2L9m} and \rf{wtildem}, while the second line is the chiral loop contribution of ref.~\cite{EPR87}, renormalized at $\nu=M_\rho$, and  where we have only retained the first two terms in the expansion of $\phi_{K}(z)$, while $\chi(z)=\phi_{\pi}(z)-\phi_{\pi}(0)$. With $\tilde {\bf w}$ and $\beta$ left as free parameters, we make a least squared fit to the experimental points in Fig.~2. The result is the continuous curve shown in the same figure, which corresponds to a $\chi_{\mbox{\rm\tiny min.}}^2=13.0$ for 18 degrees of freedom. The fitted values (using ${\bf g}_8=3.3$ and $F_{\pi}=92.4~\MeV$) are 
\begin{equation}\lbl{wtildepred}
	\tilde {\bf w}=0.045\pm 0.003\qquad\annd\qquad \beta=2.8\pm 0.1\,;
\end{equation}
and therefore
\begin{equation}\lbl{w24L9pred}
		\bfw_2 -4{\bf L}_9=-0.019\pm 0.003\,.
\end{equation}
These are the values which correspond to the cross in Fig.~1 above.

As a test we compute the $K^+\ra\pi^+ e^+ e^-$ branching ratio, using the form factor in Eq.~\rf{fff} with the fitted values for $\tilde {\bf w}$ and $\beta$, with the result
\begin{equation}
{\rm Br}(K^+\ra\pi^+ e^+ e^-)=(3.0\pm 1.1)\times 10^{-7}\,,
\end{equation}
in good agreement (as expected) with experiment result in Eq.~\rf{K+rate}.

The fitted value for $\tilde{\bf w}$ results in a negative value for $\bfw_s$ in Eq.~\rf{eq:ws} 
\begin{equation}\lbl{fitws}
	\bfw_s=-2.1\pm 0.2\,,
\end{equation}
 which corresponds to the branching ratios
 
{\setl
\bea
		{\rm Br}\left(K_S \rightarrow \pi^0 e^+ e^-\right) & = &  (7.7\pm 1.0)\times 10^{-9}\,, \\
{\rm Br}\left(K_S \rightarrow \pi^0 e^+ e^-\right)\vert_{>165{\scriptsize \MeV}} & = & (4.3\pm 0.6)\times 10^{-9} \,.
\eea}

\noi
This is to be compared with the recent NA48 results in Eq.~\rf{K0rate} and~\cite{NA48} 
\begin{equation}
{\rm Br}\left(K_S \rightarrow \pi^0 e^+ e^-\right)\vert_{>165{\scriptsize \MeV}}=		\left[3^{+1.5}_{-1.2} (\rm stat.) \pm 0.1 (\rm syst.) \right] \times 10^{-9}\,.
\end{equation}

The predicted branching ratios for the $K\ra\pi~\mu^{+}\mu^{-}$ modes are
\begin{equation}
	{\rm Br}\left(K^+ \rightarrow \pi^+ \mu^+ \mu^-\right)  =   (8.7\pm 2.8)\times 10^{-8} \quad\annd\quad	{\rm Br}\left(K_S \rightarrow \pi^0 \mu^+ \mu^-\right)  =   (1.7\pm 0.2)\times 10^{-9}\,,	
\end{equation}
to be compared with

{\setl
\bea
			{\rm Br}\left(K^+ \rightarrow \pi^+ \mu^+ \mu^-\right)  & = &  (7.6\pm 2.1)\times 10^{-8}\,, \quad {\mbox{\rm ref.~\cite{K+dacrate}}} \\
{\rm Br}\left(K_S \rightarrow \pi^0 \mu^+ \mu^-\right)  & = & \left[2.9^{+1.4}_{-1.2} (\rm stat.) \pm 0.2 (\rm syst.) \right]\times 10^{-9} \,, \quad {\mbox{\rm ref.~\cite{Moriond}}}\,.
\eea}

Finally, the resulting negative value for ${\bf w}_s$ in Eq.~\rf{fitws}, implies
 a constructive interference in Eq.~\rf{KSCPV} with a predicted branching ratio
\begin{equation}\lbl{KLCPVP}
		{\rm Br}\left(K_L \rightarrow \pi^0 e^+ e^-\right)\vert_{\rm\tiny CPV}=(3.7\pm 0.4)\times 10^{-11}\,, 
\end{equation}
 where we have used~\cite{Baetal03} $\Imm\lambda_t= (1.36\pm 0.12)\times 10^{-4}$ and we have taken into account the effect of the modulating form factor in Eq.~\rf{wtildem}.

\section{Conclusions}

Earlier analyses of $K\ra\pi~e^+ e^-$ decays within the framework of $\chi$PT have been extended beyond the predictions of $\cO(p^4)$, by replacing the local couplings which appear at that order by their underlying  narrow resonance structure in the spirit of the MHA to Large-$N_c$ QCD. The resulting modification of the $\cO(p^4)$ form factor is very simple and does not add new free parameters. It reproduces very well both the experimental decay rate and the invariant $e^+ e^-$ mass spectrum. The predicted Br$(K_S\ra\pi^0 e^+ e^-)$ and  Br$(K_S\ra\pi^0 \mu^+ \mu^-)$ are, within errors, consistent with the recently reported result from the NA48 collaboration. The predicted interference between the {\it direct} and {\it indirect} CP--violation amplitudes in $K_L\ra\pi^0 e^+ e^-$ is constructive, with an expected branching ratio (see Eq.~\rf{KLCPVP}) within reach of a dedicated experiment.

\vspace*{1cm}
{\bf Acknowledgements}

\vspace*{0.25cm}
\noi
We thank A.~Pich for his help in the earlier stages of this work and G.~Isidori for discussions. This work has been supported in part by TMR, EC-Contract No. HPRN-CT-2002-00311 (EURIDICE).

\end{document}